\documentstyle[prl,aps,preprint]{revtex}
\begin{document}

\title{Spin, gravity, and inertia}

\author{Yuri N.\ Obukhov\footnote{On leave from: Department 
of Theoretical Physics, Moscow State University, 117234 Moscow, Russia}}
\address
{Institute of Theoretical Physics, Technical University of Berlin,
Hardenbergstr. 36, D-10623 Berlin, Germany}

\maketitle

\begin{abstract}
The gravitational effects in the relativistic quantum mechanics 
are investigated. The exact Foldy-Wouthuysen transformation is 
constructed for the Dirac particle coupled to the static spacetime 
metric. As a direct application, we analyze the non-relativistic 
limit of the theory. The new term describing the specific spin 
(gravitational moment) interaction effect is recovered in the 
Hamiltonian. The comparison of the true gravitational coupling 
with the purely inertial case demonstrates that the spin relativistic 
effects do not violate the equivalence principle for the Dirac fermions. 
\end{abstract}




All high-energy physics experiments usually take place either in a curved 
spacetime or in a non-inertial reference frame (e.g., on Earth's surface 
or in the nearby space). Hence the study of the gravitational effects on  
quantum-mechanical systems represents an important issue. The weakness 
of the gravitational interaction has justified, to a certain extent, the 
long practice of neglecting the gravitational and/or inertial effects in
particle physics. 

However the technological progress, and especially the 
notable development of interferometric technique, has significantly changed 
the situation. In particular, in the famous Colella-Overhauser-Werner 
(COW) \cite{cow} and Bonse-Wroblewski \cite{bonse} experiments, the 
quantum-mechanical phase shift due to the gravitational and inertial forces 
was measured, thereby verifying the validity of the equivalence principle 
for the non-relativistic neutron waves. The corresponding theoretical 
analysis was based on the Newtonian gravity and the Schr\"odinger equation.

It is generally believed that the further improvement of the experimental
technology (using the atomic interferometers, and the polarized neutrons, 
e.g.) will soon provide a more precise picture of the interaction of quantum 
particles with the gravitational field. Under these circumstances, it seems
natural to study the higher order effects in the relativistic quantum
mechanics, including the specific manifestations of the spin-gravity coupling.

In this connection, it is worthwhile to recall that certain theoretical
models predicted the violation of the equivalence principle for spinning 
particles, see e.g. \cite{niperes,peres,mash}. A good example is provided
by the model of Peres \cite{peres}, in which the non-relativistic Hamiltonian 
of a massive Dirac particle included the additional term $k\hbar c^{-1}
\vec{\sigma}\cdot\vec{g}$. This describes the gravitational dipole type 
interaction of spin $\vec{\sigma}$ and the gravitational acceleration 
vector $\vec{g}$ with the dimensionless coupling constant $k$. Similar
interactions were considered very early by Kobzarev and Okun, and by
Leitner and Okubo \cite{gravmom}. The comparison with the precision
experimental data places though very weak restrictions on the value 
of the coupling constant $k$, see the review in \cite{mash}.

In contrast to the {\it ad hoc} Peres' type approach, here we will consider 
the standard theory of Dirac fermions in curved spacetime \cite{fock,HonnefH}. 
Correspondingly, the gravitationally coupled 4-spinor field $\psi$ 
satisfies the covariant Dirac equation:
\begin{equation}
\left(i\hbar \gamma^\alpha D_\alpha - mc\right)\psi = 0\label{Dirac0}.
\end{equation}
Here, $D_\alpha$ is the spinor covariant derivative with \cite{indices}
\begin{equation}
D_\alpha = h_\alpha^i D_i,\qquad 
D_i := \partial_i + {\frac i 4}\,\widehat{\sigma}_{\alpha\beta}
\,\Gamma_i{}^{\alpha\beta}.\label{Dspinor}
\end{equation}
We use the conventions of Bjorken and Drell \cite{bjorken} for the 
Dirac matrices $\gamma^\alpha, \beta, \vec{\alpha}$; as usual, 
$\widehat{\sigma}^{\alpha\beta}:= i\gamma^{[\alpha}\gamma^{\beta]}$.
Gravitational and inertial effects are encoded in the (co)frame and
the Lorentz connection coefficients
\begin{equation}\label{frameconn}
h_i{}^\alpha,\quad\Gamma_i{}^{\alpha\beta}= -\,\Gamma_i{}^{\beta\alpha}.
\end{equation}
As it is well known, at any point $P$, it is always possible to choose the 
local spacetime coordinates and the (non-holonomic, in general) frame so 
that $h_i{}^\alpha(P) = \delta_i^\alpha,\ \Gamma_i{}^{\alpha\beta}(P) = 0$.
This mathematical fact underlies the equivalence principle in accordance
with which the gravitationally coupled Dirac equation (\ref{Dirac0}) locally 
(at every point) assumes its flat space form in a suitably chosen reference
frame \cite{HonnefH}. 

The dynamics of the Dirac fermions in the different gravitational fields
and non-inertial reference frames was studied previously \cite{early}
using approximation schemes. Here we for the first time will present some 
{\it exact} results. More specifically, let us confine our attention to the
wide class of spacetimes described by the static metric
\begin{equation}
ds^2 = V^2\,(dx^0)^2 - W^2\,(d\vec{x}\cdot d\vec{x}),\label{metric}
\end{equation}
where $x^0 = ct$, and $V = V(\vec{x}), W = W(\vec{x})$ are the arbitrary 
functions of the spatial coordinates $\vec{x}$. Two important particular 
cases belong to this family: (i) the flat Minkowski spacetime in {\it 
accelerated frame}:
\begin{equation}
V = 1 + {\frac {(\vec{a}\cdot\vec{x})}{c^2}},\qquad W = 1,\label{accel}
\end{equation}
and (ii) {\it Schwarzschild} spacetime in isotropic coordinates:
\begin{equation}
V = \left(1 - {\frac {GM}{2c^2r}}\right)\left(1 + {\frac {GM} {2c^2r}}
\right)^{-1},\quad W = \left(1 + {\frac {GM}{2c^2r}}\right)^2,\label{schwarz}
\end{equation}
with $r := \sqrt{\vec{x}\cdot\vec{x}}$. Choosing the orthonormal frame,
\begin{equation}
h_i{}^{\widehat{0}} = V\,\delta^0_i,\qquad h_i{}^{\widehat{a}} = W
\,\delta^a_i,\qquad a,b = 1,2,3,\label{coframe}
\end{equation}
we find the local Lorentz connection:
\begin{equation}\label{connection}
\Gamma_i{}^{\widehat{a}\widehat{0}} = {\frac {\partial^a V}{WV}}
\,h_i{}^{\widehat{0}},\qquad \Gamma_i{}^{\widehat{a}\widehat{b}} 
= {\frac {\partial^aW}{W^2}}\,h_i{}^{\widehat{b}} -
{\frac {\partial^bW}{W^2}}\,h_i{}^{\widehat{a}}.
\end{equation}
Hereafter, the hats distinguish the local frame indices from the spacetime 
coordinate ones. As a result, we have the explicit spinor derivative components
\cite{sigma}:
\begin{eqnarray}
D_{\widehat{0}} &=& {\frac 1 V}\left({\frac {\partial}
{\partial x^0}} + {\frac 1 {2W}}\left(\vec{\alpha}\cdot\vec{\nabla}
V\right)\right),\label{D0}\\
D_{\widehat{a}} &=& {\frac 1 W}\left({\frac {\partial}{\partial x^a}}
+ {\frac i {2W}}\,\epsilon_{abc}\,\partial^bW\,\Sigma^c\right).\label{Da}
\end{eqnarray}
Consequently, the Dirac equation (\ref{Dirac0}) is recasted into the familiar
Schr\"odinger form
\begin{equation}\label{Dirac1}
i\hbar{\frac {\partial\psi} {\partial t}} = \widehat{\cal H}\,\psi
\end{equation}
with the Hamilton operator
\begin{eqnarray}
\widehat{\cal H} &=& \beta mc^2V + {\frac V W}\,c(\vec{\alpha}
\cdot\vec{p})\nonumber\\ 
&& - {\frac {i\hbar c}{2W}}\left(\vec{\alpha}\cdot
\vec{\nabla}V\right) - {\frac {i\hbar cV}{W^2}}\left(\vec{\alpha}
\cdot\vec{\nabla}W\right).\label{Hamilton0}
\end{eqnarray}
Redefining the spinor field and the Hamiltonian,
\begin{equation}
\psi' = W^{3/2}\,\psi,\qquad \widehat{\cal H}' = W^{3/2}
\,\widehat{\cal H}\,W^{-\,{3/2}},\label{psi1}
\end{equation}
we obtain the new Hamiltonian (which is explicitly Hermitian with respect
to the usual flat space scalar product):
\begin{equation}\label{Hamilton1}
\widehat{\cal H}' = \beta mc^2V + {\frac c 2}\left[(\vec{\alpha}
\cdot\vec{p}){\cal F} + {\cal F}(\vec{\alpha}\cdot\vec{p})\right],
\end{equation}
where ${\cal F} := V/W$. From now on we will drop the prime. 

In order to reveal the true physical content of the theory and to obtain
its correct interpretation, it is necessary to perform the Foldy-Wouthuysen 
(FW) transformation \cite{FW}, uncoupling the positive and the negative
energy states. The corresponding unitary operator can be easily obtained
for the free Dirac particle. But in most cases for a fermion interacting 
with an electromagnetic field, there is no exact transformation. Instead, 
the approximate scheme is used in which the odd parts of Hamiltonian are
removed order by order in powers of $(mc^2)^{-1}$ \cite{bjorken}. The same
approximations method was also applied in all the previous studies of the 
gravitational effects \cite{early}. 

However quite remarkably, one can construct the {\it exact FW transformation} 
for the Dirac particle moving in the metric (\ref{metric}) under 
consideration. The main guidelines are provided by Eriksen's approach to 
the FW transformation \cite{erik}. The key idea is to construct the unitary 
operator $U$, relating the FW-representation to the Dirac-representation 
$\psi^F = U\psi$, which satisfies the condition
\begin{equation}
U\,\widehat{\Lambda}\,U^\dagger = \beta.\label{ULU}
\end{equation}
Here
\begin{equation}
\widehat{\Lambda} = \widehat{\cal H}/\sqrt{{\widehat{\cal H}}^2} 
\end{equation}
is Pauli's \cite{pauli} sign energy operator. By definition, it is Hermitian,
unitary, and idempotent: $\widehat{\Lambda}^2 = \widehat{\Lambda}^\dagger
\widehat{\Lambda} = 1$ (as usually, one should assume that Hamiltonian is
well defined in that it does not possess zero eigenvalues, see \cite{erik}). 

It is now crucial to observe that the Hamiltonian (\ref{Hamilton1}) 
admits the anticommuting involution operator
\begin{equation}
J := i\gamma_5\beta.\label{invol}
\end{equation}
Clearly, this operator is Hermitian, $J^\dagger = J$, and unitary, 
$JJ^\dagger = J^2 = 1$. It anticommutes both with the Hamiltonian
(\ref{Hamilton1}) and with the $\beta$ matrix:
\begin{equation}
J\widehat{\cal H} + \widehat{\cal H}J = 0,\qquad 
J\beta + \beta J = 0.\label{involcom}
\end{equation}
Now, it is straightforward to see that the FW transformation is realized 
by means of the operator $U = U_2\,U_1$, where
\begin{equation}
U_1 = {\frac 1 {\sqrt{2}}}\left(1 + J\,\widehat{\Lambda}\right),
\qquad U_2 = {\frac 1 {\sqrt{2}}}\left(1 + \beta J\right).\label{U12}
\end{equation}
Indeed, we immediately find
\begin{equation}
U_1\widehat{\Lambda}U_1^\dagger = J,\qquad
U_2 J U_2^\dagger = \beta,\label{ULUJ}
\end{equation}
and consequently (\ref{U12}) satisfies the FW condition (\ref{ULU}). 

The final step is to find the Hamiltonian $\widehat{\cal H}^F = U
\widehat{\cal H}U^\dagger$ in the FW-representation. From (\ref{involcom})
we have $J\widehat{\cal H}^2 = \widehat{\cal H}^2J$, hence $J\sqrt{
{\widehat{\cal H}}^2} = \sqrt{{\widehat{\cal H}}^2}J$, and $J\widehat{\Lambda} 
+ \widehat{\Lambda}J = 0$. Consequently, one finds that 
$U_1\widehat{\cal H}U^\dagger_1 
= J\sqrt{{\widehat{\cal H}}^2}$, and finally, 
\begin{equation}
U\widehat{\cal H}U^\dagger = U_2\,U_1\,\widehat{\cal H}
\,U_1^\dagger\,U_2^\dagger = \left[\sqrt{{\widehat{\cal H}}^2}\;\right]
\!\beta + \left\{\sqrt{{\widehat{\cal H}}^2}\right\}\!J.\label{HFW}
\end{equation}
Here, as usually, the even and odd parts of any operator $Q$ are defined as
\begin{equation}
[Q] := {\frac 1 2}\left(Q + \beta Q\beta\right),\qquad 
\{Q\} := {\frac 1 2}\left(Q - \beta Q\beta\right).\label{evenodd}
\end{equation}
Note that both terms in (21) are clearly even, thus indeed the FW Hamiltonian
does not mix the upper and lower spinor components (i.e., positive and negative
energy states). 

Usually (at least in the known cases when the exact FW transformation
exists for the electromagnetic coupling), the square of Hamiltonian turns
out to be an even operator, and then the second term in (\ref{HFW}) is
absent. However, this is not the case for our problem, because the 
square of Hamilton operator 
\begin{eqnarray}
\widehat{\cal H}^2 &=& m^2c^4\,V^2 + {\cal F}c^2p^2{\cal F} + 
{\frac {\hbar^2c^2}2}\,{\cal F}(\vec{\nabla}\cdot\vec{f}) 
- {\frac {\hbar^2c^2} 4}\vec{f}^2 \nonumber\\
&& +\,\hbar c^2\,{\cal F}\,\vec{\Sigma}\cdot
\left([\vec{f}\times\vec{p}\,] + J\,mc\,\vec{\phi}\,\right)\label{Ham2}
\end{eqnarray}
contains the {\it odd} piece (the very last term in the above expression).
Here we denoted the gradients
\begin{equation}\label{grad}
\vec{\phi} := \vec{\nabla}V,\qquad\vec{f}:= \vec{\nabla}{\cal F}.
\end{equation}

The FW Hamiltonian (\ref{HFW}) is {\it exact}. Now we may turn to its
analysis, looking for different important limiting cases. For the most
practical purposes, it is sufficient to use the non-relativistic wave
functions, treating all the interaction terms as perturbations. The 
quasi-relativistic approximation is straightforwardly obtained by
assuming that $mc^2$ term is dominating, and thus correspondingly 
expanding the square root of (\ref{Ham2}) as
\begin{eqnarray}
\sqrt{\widehat{\cal H}^2} &\approx& mc^2\,V + {\frac 1 {4m}}
\left(W^{-1}p^2{\cal F} + {\cal F}p^2W^{-1}\right)
+ {\frac {\hbar^2}{4mW}}\,(\vec{\nabla}\cdot\vec{f}) 
- {\frac {\hbar^2} {8mV}}\vec{f}^2 \nonumber\\
&& +\,{\frac {\hbar} {4m}}\,\vec{\Sigma}\cdot\left(W^{-1}
\,[\vec{f}\times\vec{p}\,] + [\vec{f}\times\vec{p}\,]\,W^{-1}
+ J\,2W^{-1}mc\,\vec{\phi}\,\right),\label{sqrtHap}
\end{eqnarray}
with the subsequent extraction of the even and odd pieces according to 
(\ref{evenodd}). It seems worthwhile to note the appearance of the 
``gravitational Darwin'' term
\begin{equation}
{\frac {\hbar^2}{4mW}}\,(\vec{\nabla}\cdot\vec{f}) =
{\frac {\hbar^2}{4mW}}\,\Delta{\cal F},\label{darwin}
\end{equation}
which was not reported before \cite{early}. It clearly admits a physical 
interpretation similar to that of the usual electromagnetic Darwin term,
reflecting the zitterbewegung fluctuation of the fermion's position with
the mean square $<\!\!(\delta r)^2\!\!>\sim\hbar^2/(mc)^2$. Analyzing the 
effective gravitostatic energy instead of the electrostatic one (with 
the charge $e$ replaced by the mass $m$), one then can derive a contribution
of the form (\ref{darwin}).

In order to compare the relativistic spin effects of the gravitational 
and inertial forces, and thereby to obtain an insight into the validity
of the equivalence principle for fermions, it is instructive to 
consider separately the two above mentioned particular cases (\ref{accel}) 
and (\ref{schwarz}) of the metric (\ref{metric}).

{\it (i) Accelerated frame}. 
{}From (\ref{accel}) we find ${\cal F} = V$, and consequently:
\begin{equation}
\vec{\phi} = \vec{f} = {\frac {\vec{a}} {c^2}}.\label{gradA}
\end{equation}
Preserving the leading contributions, we then find for the non-relativistic
FW Hamiltonian:
\begin{eqnarray}
\widehat{\cal H}^F &=& \beta mc^2 + \beta m\vec{a}\cdot\vec{x} + \beta\,
{\frac {p^2}{2m}} + {\frac \hbar {2c}}\,\vec{\Sigma}\cdot\vec{a}\nonumber\\ 
&& +\,{\frac \hbar {2mc^2}}\,\beta\vec{\Sigma}\cdot[\vec{a}\times
\vec{p}].\label{Hacc}
\end{eqnarray}
The Darwin term identically vanishes for obvious reasons. 

{\it (ii) Spherically symmetric gravitational field}. Far away from the 
central gravitating body of a mass $M$, it is sufficient to consider a 
weak limit of the Schwarzschild field (\ref{schwarz}) which yields
\begin{equation}
V \approx 1 - {\frac {GM} {c^2r}},\qquad W \approx 1 + {\frac {GM} {c^2r}}.
\end{equation}
Correspondingly, 
\begin{equation}
\vec{\phi} = -\,{\frac {\vec{g}} {c^2}},\qquad \vec{f} 
= -\,{\frac {2\vec{g}} {c^2}}, \quad {\rm with}\quad 
\vec{g} = -\,GM {\frac {\vec{r}} {r^3}},\label{gradS}
\end{equation}
and the non-relativistic FW Hamiltonian reads: 
\begin{eqnarray}
\widehat{\cal H}^F &=& \beta mc^2 + \beta m\vec{g}\cdot\vec{x} + \beta\,
{\frac {p^2} {2m}} - {\frac {\hbar}{2c}}\,\vec{\Sigma}\cdot\vec{g}\nonumber\\
&& -\,{\frac {\hbar} {mc^2}}\,\beta\vec{\Sigma}\cdot[\vec{g}\times\vec{p}]
- {\frac {\hbar^2\beta}{2m}}\,(\vec{\nabla}\cdot\vec{g}).\label{Hsch}
\end{eqnarray}
In both cases we neglect in (\ref{Hacc}) and (\ref{Hsch}) the higher order
relativistic and gravitational/inertial (``red shift'' etc) terms. 

The most nontrivial common feature is the recovery of the Peres' type
spin contribution [last terms on the first lines in (\ref{Hacc}) and 
(\ref{Hsch})]. We have however two important differences from \cite{peres}. 
Firstly, unlike the ad hoc model of Peres, the corresponding coupling 
constant is now fixed: $k = 1/2$. Secondly, the same spin term is 
present in both cases, 
one just needs to replace the acceleration $\vec{a}$ of a 
reference system by the true gravitational acceleration $\vec{g}$. Hence, 
in contrast to Peres' model, the covariant Dirac theory proves the validity 
of the equivalence principle, also with the higher order relativistic spin 
effects taken into account. 

One can estimate (following \cite{peres}) the influence of the 
additional term. For the gravitational field of the Earth one has 
$g/c = 3.271\times 10^{-8}$Hz, or $\hbar g/c$ 
$=2.153\times 10^{-23}$eV, which is essentially smaller than what the 
present experimental technique can detect [the next terms in (\ref{Hacc}) 
and (\ref{Hsch}) are several orders weaker].  
At the same time, this is also in a good agreement with the 
strong limits set by the high precision measurements, e.g., of the 
hyperfine splitting, \cite{gravmom}. 
A possible direct test might be carried out with a spin-polarized 
macroscopic body \cite{peres1}.

A comment is needed for explaining the difference of our 
derivations with the earlier results \cite{early}. On the one hand, it is 
important to realize that the approximate scheme \cite{bjorken} developed 
for the case of electromagnetic coupling is not, strictly speaking, 
applicable in the gravitational interaction case. The idea of such an 
approximate scheme is to remove, order by order in $1/m$, the odd terms 
from the Hamiltonian $\widehat{\cal H} = \beta mc^2 + {\cal E} + {\cal O}$. 
Normally, the odd ${\cal O}$ and even ${\cal E}$ parts did not depend on 
the mass $m$: All the terms were proportional to the {\it electromagnetic} 
charge $e$, and that made the standard scheme \cite{bjorken} working. 
However, for the gravitational/inertial case, the even part ${\cal E}$ 
necessarily contains the terms proportional to the {\it 
gravitational/inertial} charge $m$. As a result, although in the first 
approximation the original odd term ${\cal O}$ is removed, the new term 
is produced $\beta [{\cal O},{\cal E}]/(2m)$ which gives a contribution 
of order $m^0$. The same is repeated at every step of the approximate 
scheme, so that the remaining even terms are always of the same order 
in $1/m$ as the ``removed'' odd terms. This makes the crucial issue of 
convergence of the approximation scheme problematic \cite{costella}. In 
our approach, we avoid this deficiency by using the {\it exact} FW 
transformation. It seems worthwhile to notice that the method works 
also for the case with the magnetic field coupling included, when a 
generalization of the result of Case \cite{case} is obtained. On the 
other hand, the FW transformation is defined with a certain ambiguity. 
The unitary transformation $U=e^{iS}$ with $S = - \beta/(4mc)\vec{\Sigma}
\cdot\left(W^{-1}\vec{p} + \vec{p}\,W^{-1}\right)$ brings the Hamiltonian 
$\widehat{\cal H}^F\rightarrow U\widehat{\cal H}^FU^\dagger$ to the 
approximate form reported by Fischbach et al and by Hehl and Ni \cite{early}.

A spin carried by a fermion can be visualized (with the obvious 
reservations and a reasonable portion of caution \cite{yang}) as some 
sort of intrinsic circular motion. In this primitive picture, a particle's 
electric charge induces an Amp\`ere type ring current which, in turn, 
according to the Oersted-Amp\`ere law, acts as a magnetic moment. 
The massive fermion, besides the electric charge, carries also the 
{\it gravitational charge}. Accordingly, in the framework of general 
relativity, one can naturally expect a mass-energy ring current inducing 
a gravitational moment \cite{sch}. In this paper we have demonstrated how 
the gravitational moment can show up explicitly in the non-relativistic
limit of the covariant Dirac theory. 

\bigskip
{\bf Acknowledgments}. This work was supported by the Deutsche 
Forschungsgemeinschaft (Bonn).



\begin{references}                    
\bibitem{cow}
R. Colella, A.W. Overhauser, and S.A. Werner, 
{\sl Phys. Rev. Lett.} {\bf 34} (1975) 1472. 

\bibitem{bonse}
U. Bonse and T. Wroblewski, 
{\sl Phys. Rev. Lett.} {\bf 51} (1983) 1401. 

\bibitem{niperes}
T.A. Morgan and A. Peres, 
{\sl Phys. Rev. Lett.} {\bf 9} (1962) 79; 
W.-T. Ni, 
{\sl Phys. Rev. Lett.} {\bf 38} (1977) 301; 
N.D. Hari Dass, 
{\sl Phys. Rev. Lett.} {\bf 36} (1976) 393; 
N.D. Hari Dass, 
{\sl Ann. Phys. (NY)} {\bf 107} (1977) 337. 

\bibitem{peres}
A. Peres, 
{\sl Phys. Rev.} {\bf D18} (1978) 2739. 

\bibitem{mash}
B. Mashhoon, 
{\sl Class. Quantum Grav.} {\bf 17} (2000) 2399. 

\bibitem{gravmom}
I.Yu. Kobzarev and L.B. Okun, 
{\sl Sov. Phys. JETP} {\bf 16} (1963) 1343 
[{\sl ZhETF} {\bf 43} (1962) 1904 
(in Russian)];
J. Leitner and S. Okubo, 
{\sl Phys. Rev.} {\bf 136} (1964) 1542. 

\bibitem{fock}
V.A. Fock and D.D. Ivanenko, 
{\sl C.R. Acad. Sci. Paris} {\bf 188} (1929) 1470; 
V.A. Fock and D.D. Ivanenko, 
{\sl Z. Phys.} {\bf 54} (1929) 798; 
H. Tetrode, 
{\sl Z. Phys.} {\bf 50} (1928) 336;
V. Bargmann, 
{\sl Sitzungsber. preuss. Akad. Wiss. Phys.-math. Kl.} (1932) 346;
E. Schr\"odinger E., 
{\sl Sitzungsber. preuss. Akad. Wiss. Phys.-math. Kl.} (1932) 105; 
L. Infeld L. and B.L. van der Waerden, 
{\sl Sitzungsber. preuss. Akad. Wiss. Phys.-math. Kl.} (1933) 380; 474;
D.R. Brill and J.A. Wheeler, 
{\sl Rev. Mod. Phys.} {\bf 29} (1957) 465, 
{\it Errata}: {\sl Rev. Mod. Phys.} {\bf 33} (1961) 623.

\bibitem{HonnefH}
See also a general discussion in: F.W. Hehl, J. Lemke, and E.W. Mielke, 
{\it Two lectures on fermions and gravity}, in: {\sl Geometry and 
Theoretical Physics}, Proc. of the Bad Honnef School 12--16 Feb.\ 
1990, J. Debrus and A.C. Hirshfeld, eds.\ (Springer: Heidelberg, 1991) 
pp.\ 56--140.

\bibitem{indices}
We use the Greek alphabet for the indices which label the components
with respect to a local Lorentz frame $e_\alpha = h_\alpha^i\partial_i$,
whereas the Latin indices refer to the local spacetime coordinates $x^i$. 

\bibitem{bjorken}
J.D. Bjorken and S.D. Drell, {\it Relativistic Quantum Mechanics}
(McGraw-Hill: San Francisco, 1964).

\bibitem{early}
C.G. de Oliveira and J. Tiomno, 
{\sl Nuovo Cim.} {\bf 24} (1962) 672; 
J. Audretsch and G. Sch\"afer, 
{\sl Gen. Rel. Grav.} {\bf 9} (1978) 243; 
E. Fischbach, B.S. Freeman, and W.-K. Cheng, 
{\sl Phys. Rev.} {\bf D23} (1981) 2157; 
F.W. Hehl and W.-T. Ni, 
{\sl Phys. Rev.} {\bf D42} (1990) 2045; 
Y.Q. Cai and G. Papini, 
{\sl Phys. Rev. Lett.} {\bf 66} (1991) 1259; 
{\bf 68} (1992) 3811;
J.C. Huang, {\sl Ann. d. Phys.} {\bf 3} (1994) 53;
K. Konno and M. Kasai, 
{\sl Prog. Theor. Phys.} {\bf 100} (1998) 1145; 
K. Varj\'u and L.H. Ryder, 
{\sl Phys. Lett.} {\bf A250} (1998) 263; 
L. Ryder, 
{\sl J. Phys. A: Math. Gen.} {\bf A31} (1998) 2465. 
K. Varj\'u and L.H. Ryder, 
{\sl Phys. Rev.} {\bf D62} (2000) 024016. 

\bibitem{sigma}
As usually, we have $\beta = \gamma^{\widehat{0}}, \vec{\alpha} = \beta
\vec{\gamma}$, $\gamma_5 = -i\gamma^{\widehat{0}}\gamma^{\widehat{1}}
\gamma^{\widehat{2}}\gamma^{\widehat{3}}$. The {\it spin matrix} is 
$\vec{\Sigma} = i\vec{\gamma}\times\vec{\gamma}/2 = -\gamma_5\vec{\alpha} = 
\left(\begin{array}{cc} \vec{\sigma}&0\\ 0&\vec{\sigma}\end{array}\right)$.

\bibitem{FW}
L.L. Foldy and S.A. Wouthuysen, 
{\sl Phys. Rev.} {\bf 78} (1950) 29. 

\bibitem{erik}
E. Eriksen and M. Kolsrud, 
{\sl Nuovo Cim.} {\bf 18} (1960) 1; 
A.G. Nikitin, 
{\sl J. Phys. A: Math. Gen.} {\bf A31} (1998) 3297. 

\bibitem{pauli}
W.~Pauli, {\it Die allgemeinen Prinzipien der Wellenmechanik}, in: 
{\sl Handbuch der Physik}, Ed. S.~Fl\"ugge (3.~Aufl., Berlin, 1958) 
Bd. 5, Teil 1, S. 1-168.

\bibitem{peres1}
An idea of one such experiment, as proposed in \cite{peres}, is to
observe a breaking of the equilibrium of a polarized body, hanging
in the gravitational field, when its polarized state is destroyed. The 
main difficulty here is to provide a very high degree of shielding
of external magnetic fields.

\bibitem{case}
K.M. Case, 
{\sl Phys. Rev.} {\bf 95} (1954) 1323. 

\bibitem{costella}
The danger of a failure of convergence in presence of terms of positive 
orders in $m$ is discussed, e.g., by J.P. Costella and B.H.J. McKellar, 
{\sl Am. J. Phys.} {\bf 63} (1995) 1119. 

\bibitem{yang}
See, though, the careful analysis by 
T.T. Chou and C.N. Yang, 
{\sl Nucl. Phys.} {\bf B107} (1976) 1, 
who conclude that ``a polarized Dirac electron is a 
{\it rotating} particle'' which carries the electric current.

\bibitem{sch}
The notion of the gravitational moment was proposed in \cite{gravmom}.
For the recent studies, see 
F.W. Hehl, A. Macias, E.W. Mielke, and Yu.N. Obukhov, 
in: {\sl ``On Einstein's path" Festschrift for E.~Schucking}. 
A. Harvey, ed. (Springer: New York, 1999) 255-274;
Yu.N. Obukhov, 
{\sl Acta Phys. Polon.} {\bf B29} (1998) 1131. 


\end{references}
\end{document}